\documentclass[11pt,twoside]{article} 

\usepackage{epsf}

\usepackage{amsmath}
\usepackage{amssymb}
\usepackage{psfig}
\usepackage{latexsym}
\usepackage{graphicx}
\usepackage{epic}
\usepackage{eepic}
\usepackage{setspace}
\usepackage{supertabular}
\usepackage{psfrag}
\usepackage{rotating}
\usepackage{wrapfig}
\usepackage{fancyheadings}
\usepackage[dvips]{color}
\usepackage{cite}
\usepackage{makeidx}
\usepackage[english]{babel}
 \usepackage{eufrak}
\usepackage{bm}

\newcommand{\beq}{\begin{equation}}
\newcommand{\eeq}{\end{equation}}
\newcommand{\ket}[1]{|#1\rangle}
\newcommand{\bra}[1]{\langle #1|}
\newcommand{\UCU}[2]{F^{#1}_{#2}\left(U\!\left(2\right)\right)}
\newcommand{\UC}[3]{F^{#1}_{#2}\left( #3 \right)}
\newcommand{\TUCU}[2]{\tilde{F}^{#1}_{#2}\left(U\!\left(2\right)\right)}

\date{}

\title{Decompositions of general quantum gates}  
\author{Mikko M\"ott\"onen and Juha J. Vartiainen} 

\begin{document}

\maketitle     

\begin{center}\emph{Materials Physics Laboratory, POB 2200 (Technical Physics)}\\
\emph{FIN-02015 HUT, Helsinki University of Technology, Finland}\end{center}

\section*{Abstract}
Quantum algorithms may be described by sequences of unitary transformations called quantum gates and measurements applied to the quantum register of $n$ quantum bits, qubits. A collection of quantum gates is called universal if it can be used to construct any $n$-qubit gate. In 1995, the universality of the set of one-qubit gates and controlled~{\footnotesize{NOT}} gate was shown by Barenco {\it et al.} using QR decomposition of unitary matrices. Almost ten years later the decomposition was improved to include essentially fewer elementary gates. In addition, the cosine-sine matrix decomposition was applied to efficiently implement decompositions of general quantum gates. In this chapter, we review the different types of general gate decompositions and slightly improve the best known gate count for the controlled~{\footnotesize{NOT}} gates to $\frac{23}{48}4^n$ in the leading order. In physical realizations, the interaction strength between the qubits can decrease strongly as a function of their distance. Therefore, we also discuss decompositions with the restriction to nearest-neighbor interactions in a linear chain of qubits.

\section{Introduction}\label{intro}
The emerging of quantum mechanics~\cite{Ballentine} in the beginning of the 20$^{{\rm th}}$ century revolutionized the field of physics bringing not only understanding to fundamental concepts such as atomic and particle physics, but also numerous applications for everyday life. One of the most important applications are the semiconductors, namely the transistor which is the basis of today's digital computers. As quantum mechanics shook up physics, quantum computing~\cite{NielsenChuang} has done the same for computer science. Some quantum algorithms, Shor's algorithm~\cite{shor94} being the most famous, offer exponential speedup compared with the best known classical counterparts due to the phenomenon called quantum parallelism. Shor's algorithm may be used to break the commonly used RSA encryption for key distribution but, on the other hand,  quantum physics also provides a secure information channel using quantum cryptography~\cite{gisin}. Due to its powerful applications, the experimental realization of the quantum computer is regarded as highly important issue in physics. Similarly, theoretical research which lightens the burden of the experimental needs is also of great interest.

In quantum computing, the algorithms are commonly described by the quantum circuit model~\cite{deutsch}. It involves quantum gates, projective measurements and a register of $n$ quantum bits, called qubits. In a classical computer, a bit may have only two distinct values usually denoted by 0 and 1. In contrast, a qubit may be in a superposition of these two basis vectors, {\it i.e.}, the state of the qubit is described by a vector in the complex space $\mathbb{C}^2$. In this space, the quantum gates correspond to matrices, which are unitary due to the unitary temporal evolution of any closed quantum system. 

Since many algorithms involve gates acting on $n$ qubits, it is an important issue how these gates may be decomposed into an array of simpler gates accessible to the experiments. In general, we may assume that we have a collection of simple quantum gates, called the gate library, into which the $n$-qubit gates are to be decomposed. The gates in the gate library are called elementary gates. The library is called universal if any $n$-qubit gate has a presentation only involving gates from that library. We choose our library to consist of all one-qubit gates and the controlled~{\footnotesize{NOT}} gate ({\footnotesize{CNOT}}) which are defined in Sec.~\ref{circuits}. This particular library has been proved to be universal~\cite{barenco} but, actually, almost any other two-qubit gate could be chosen to replace the {\footnotesize{CNOT}} for the universality to hold~\cite{lloyd}. However, it is feasible to work with the {\footnotesize{CNOT}} since it has a rather simple logical structure. 

The proof of the universality of our gate library~\cite{barenco} was, in fact, constructive but the number of {\footnotesize{CNOT}}s involved was as high as $O(n^34^n)$. It is convenient to calculate the number of {\footnotesize{CNOT}}s and one-qubit gates separately, since {\footnotesize{CNOT}}s introduce interactions between the qubits and those interactions are usually much weaker than the interactions between a single qubit and the control fields. Hence, the experimental realization of the {\footnotesize{CNOT}} is typically a much slower process than that of a one-qubit gate. Already in 1995, it was shown that the circuit complexity could be reduced down to $O(n4^n)$~\cite{knill}, but until the year 2004 there was no remarkable progress on the decomposition of arbitrary quantum gates. Reference~\cite{cybenko} reviews briefly the traditional decomposition of Ref.~\cite{barenco}.

The highest known lower bound for the number of {\footnotesize{CNOT}}s needed to decompose a general unstructured quantum gate acting on $n$ qubits is $\lceil (4^n-3n-1)/4\rceil$~\cite{shende} and, hence, there was an extra factor of $n$ in the best known complexity. Finally in 2004, being an unsolved mystery for about ten years the original construction was improved to yield the complexity $O(4^n)$~\cite{QR_PRL}. However, this decomposition was still far from the lower bound. The original gate decomposition made use of the QR matrix decomposition~\cite{golub}. In contrast, Ref.~\cite{CSD_PRL} introduces the cosine-sine matrix decomposition\footnote{In context of quantum computing, the CSD was discussed first in Ref.~\cite{tucci}.} (CSD)~\cite{golub} in this context which turned out to yield a leading order complexity $4^n$ for the one-qubit gates and $4^n$ for the {\footnotesize{CNOT}}s. The CSD was also combined with a so-called quantum multiplexor (QM), a special method to simplify the gates, to obtain a decomposition involving $4^n/2$ {\footnotesize{CNOT}}s and the same number of one-qubit gates in the leading order~\cite{pitkapaperi} (see also Ref.~\cite{shende_matrix}). In this chapter, we present an improvement to the decomposition introduced in Ref.~\cite{shende_matrix} to obtain the lowest {\footnotesize{CNOT}} count known to date.

This chapter is organized as follows. In Sec.~\ref{notation}, we define our notation and introduce some of the important mathematical tools. Section~\ref{ucr} is devoted to the presentation of so-called uniformly controlled gates (UCGs) and their efficient decomposition into elementary gates. The UCGs are the natural building blocks of decompositions employing the CSD. The original QR decomposition and its improved versions are discussed in Sec.~\ref{qr} in contrast to Sec.~\ref{csd} in which the CSD is studied. Finally, the local state preparation, {\it i.e.}, the question how to transform any given quantum state into another arbitrary state, is implemented Sec.~\ref{lsp} following Refs.~\cite{stateprep,shende_matrix,pitkapaperi}. The state preparation may be useful if one wishes to use, for example, exotic inputs to algorithms. In Sec.~\ref{conclusion}, we conclude and summarize our discussions.

\section{Preliminaries}\label{notation}

\subsection{Quantum state and unitary temporal evolution}\label{qsute}
We consider here a quantum register consisting of $n$ qubits and, hence, all possible quantum states of the system are in the Hilbert space $\mathcal{H}:=\bigotimes_{i=1}^{n}\mathbb{C}^2=\mathbb{C}^{2^n}$\!, where the symbol $\otimes$ denotes the Kronecker product. The basis vectors for each of the qubits are chosen as
\beq
\ket{0}=\left(%
\begin{array}{c}
  1 \\
  0 \\
\end{array}%
\right)
\quad {\rm and} \quad
\ket{1}=\left(%
\begin{array}{c}
  0 \\
  1 \\
\end{array}%
\right).
\eeq
For the whole configuration space $\mathcal{H}$, it is convenient to choose the basis vectors to be $\{\ket{e_k}\}$, $k=1,\ldots,N:=2^n$. Here $|e_k\rangle=\bigotimes_i |x_i^k\rangle$, where
$x_i^k \in \{0,1\}$ and the index $i=1,...,n$ refers to the qubit. 
In this basis the state vector of the system is of the form
\beq
\label{eq:superpos}
\ket{\Psi}=\sum_{i=1}^{N} a_i \ket{e_i}\, \quad {\rm and} \quad \sum_{i=1}^N|a_i|^2=1,
\eeq
where the latter equality fixes the normalization of the vector. Hence, the probability for the system to be in a state $\ket{e_i}$ after a projective measurement is  $|a_i|^2$. It is also noted that the global phase of the state vector is unobservable and, hence, may be taken to unity\footnote{Clearly the global phase does not affect the probabilities. Furthermore, addition of a global phase commutes with any unitary matrix, {\it i.e.}, it has no effect on the temporal evolution of the system.}.

Conventionally in quantum computing, the order of the basis vectors
has been chosen such that the values $x_i^k$ essentially form the
binary representation of the number $k-1$, {\it i.e.}, $k=1+\sum_{i=0}^n
2^i x_i^k$. We note that the order of the basis vectors in the
computational basis can be freely chosen. We will make use of this degree of freedom in Sec.~\ref{qr} in the context of the QR decomposition.

The fundamental differences of the quantum computer compared with
the classical one arise from the utilization of the high-dimensional Hilbert space $\mathcal{H}$.
In comparison, the states accessible to a classical computer are limited to the basis vectors
$\ket{\Psi}=\ket{e_i}$, {\it i.e.}, to the states in which all of the weight factors except one vanish.
The quantum mechanical superposition principle allows several
weight factors to be simultaneously non-zero, which renders the quantum mechanical state space greatly larger than the classical one.

The temporal evolution of any quantum system is governed by the well known Schr\"{o}dinger equation
\beq
\label{shrode}
i \hbar \frac{\partial}{\partial t}\ket{\Phi(t)} = {H}\ket{\Phi(t)},
\eeq
where the Hamiltonian ${H}$ of the pure quantum system is always Hermitian. This implies that the temporal evolution may be described by a unitary operator $\mathcal{U}(t,0)$ as $\ket{\Phi(t)}=\mathcal{U}(t,0)\ket{\Phi(0)}$. In our finite dimensional Hilbert space the unitary operator may be written as a unitary matrix $U\in SU(N)$. The reason why the determinant of $U$ may be taken to unity is that the global phase of the state vector has no physical meaning. Since the $n$-qubit quantum gate may be represented by a unitary matrix, it is reasonable that the gate decompositions may correspond to some known matrix decompositions and vice versa. 

\subsection{Quantum circuits}\label{circuits}
A one-qubit gate $U \in SU(2)$ acting on the $k^{\rm th}$ qubit in a $n$-qubit register is represented by a unitary matrix
\beq
\tilde{U}=\underbrace{I \otimes \ldots \otimes I}_{k-1\,\,{\rm times}} \otimes U \underbrace{\otimes I  \ldots \otimes I}_{n-k\,\,{\rm times}},
\eeq
For simplicity,
we omit below the qubits that are operated on only by an identity operator.
Accordingly, the matrix representation of the gate $U$ is
\begin{equation}
U=\left(%
\begin{array}{cc}
  a & b \\
  -\overline{b} & \overline{a} \\
\end{array}%
\right),
\end{equation}
where $a$ and $b$ are two complex numbers satisfying $|a|^2+|b|^2=1$.
We fix the basis for the two-state system such that the generator $\sigma_z$ of the $SU(2)$ group is diagonal. Furthermore, we
call the vectors corresponding to the eigenvalues 1 and -1 by $\ket{0}$ and  $\ket{1}$, respectively.
In this basis the matrix representations of generators $\{ \sigma_i \}$ are called the Pauli spin matrices
\begin{equation} \label{eq:pauli}
\sigma_x=
\begin{pmatrix}
  0 & 1 \\
  1 & 0 \\
\end{pmatrix},
\quad
\sigma_y=
\begin{pmatrix}
  0 & -i \\
  i & 0 \\
\end{pmatrix},
\quad
\sigma_z=
\begin{pmatrix}
  1 & 0 \\
  0 & -1 \\
\end{pmatrix}.
\end{equation}
In fact, any $U\in SU(2)$ may be written as a rotation 
\begin{equation}\label{rota}
U=R_{{\bf a}}(\theta) = e^{i  {\bf a} \cdot \bm{\sigma}\theta/2} =
I \cos \frac{\theta}{2} + i \left( {\bf a} \cdot \bm{\sigma}
\right)\sin \frac{\theta}{2},
\end{equation}
where the symbol $\theta$ stands for the rotation angle around the unit vector ${\bf a}$ fixed by $U$ and we have introduced the product ${\bf a} \cdot\bm{\sigma}=a_x\sigma_x+a_y\sigma_y+a_z\sigma_z$.
Equation~(\ref{rota}) yields that any rotation $R_{{\bf a}}(\theta)$ can be made diagonal as
\beq
R_{{\bf a}}(\theta)= V_{{\bf a}}R_z(\theta) V_{{\bf a}}^\dagger,
\label{eq:diag}
\eeq
where the similarity transformation $V_{{\bf a}}$ diagonalizes the matrix ${\bf a} \cdot \bm{\sigma}$.
We note that the matrix $V_{{\bf a}}$ does not depend on the rotation angle $\theta$.
In addition, all rotations about any single axis are additive
\begin{equation}
R_{{\bf a}}(\theta_1) R_{{\bf a}}(\theta_2) = R_{{\bf a}}(\theta_1
+ \theta_2), \label{eq:mopo}
\end{equation}
and the rotation angle of all rotations with $a_x = 0$ is reversed by conjugation with $\sigma_x$ as
\begin{equation}
a_x = 0 \implies \sigma_x R_{{\bf a}}(\theta) \sigma_x = R_{{\bf
a}}(-\theta). \label{eq:mopo2}
\end{equation}
The rotations for which the rotation vector is parallel to any of the coordinate axes are called elementary rotations and denoted by $R_x(\theta)$, $R_y(\theta)$ and $R_z(\theta)$. Any element $U\in SU(2)$ may be written using only two different types of elementary rotations, {\it e.g.}, $z$ and $y$ rotations as
\beq
U=R_z(\alpha)R_y(\beta)R_z(\gamma),
\eeq
where angles~$\alpha, \beta, \gamma$ are called the Euler angles. The above results are used in the next sections to achieve and simplify the studied gate decompositions.

\begin{figure}
\center
\includegraphics[width=0.65\textwidth]{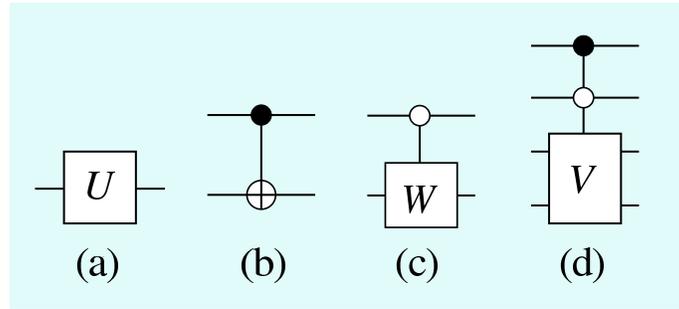}
\caption{\label{fig1} Quantum circuit symbols for (a) one-qubit gate, (b) {\footnotesize{CNOT}}, (c) controlled
one-qubit gate, (d) twofold controlled two-qubit gate. In (c) the gate $W$ acts only on the subspace in which the control qubit lies in the state $\ket{0}$ and in (d) the gate $V$ operates on subspace in which the control qubits are in the state $\ket{10}$.}
\end{figure}

The circuit diagram for the one-qubit gate $U$ is shown in Fig.~\ref{fig1}(a). The only two-qubit gate in out library is the {\footnotesize{CNOT}} shown in Fig.~\ref{fig1}(b). The action of the {\footnotesize{CNOT}} is logical {\footnotesize NOT} in the subspace $\{\ket{10}, \ket{11}\}$ and it leaves the subspace where the value of the control qubit (the upper qubit) is zero untouched. The matrix presentation for the {\footnotesize{CNOT}}
in basis $\{\ket{00},\ket{01},\ket{10},\ket{11} \}$ is
\begin{equation}
U_{\rm CNOT} = I \oplus \sigma_x =
\left(%
\begin{array}{cccc}
  1 & 0 & 0 & 0 \\
  0 & 1 & 0 & 0 \\
  0 & 0 & 0 & 1 \\
  0 & 0 & 1 & 0 \\
\end{array}%
\right).
\end{equation}
In general, the qubits are denoted by horizontal lines in the quantum circuit diagrams and the gates as rectangles. The control nodes are marked by circles which are connected to the associated gate by a vertical line. The effect of the control nodes is to limit the corresponding gates to act only on the subspace characterized by its control nodes. The nodes in the quantum circuit diagram can be black or white corresponding to the control qubit states $\ket{1}$  or $\ket{0}$, respectively (see Figs.~\ref{fig1}(c) and~(d)). Hereafter we refer to the $k$-fold controlled one-qubit gate $V$ by C$^{k}V$. When applied to an $n$ qubit register, this gate operates in $2^{n-k}$-dimensional target subspace consisting of those basis vectors for which the values of the control qubits match with those of the control nodes.

\section{Uniformly controlled gates}\label{ucr}

\subsection{Decomposition of uniformly controlled elementary rotations}\label{ucer}
\begin{figure}
\center
\includegraphics[width=0.75\textwidth]{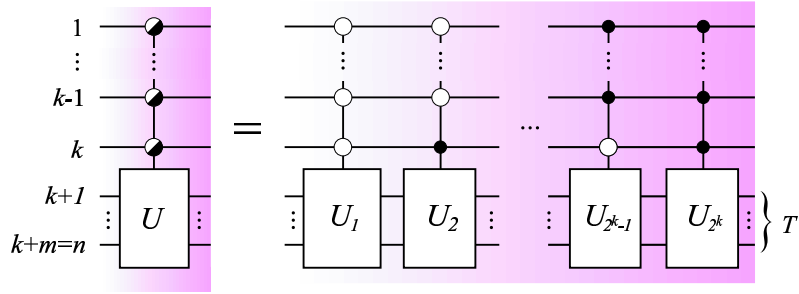}
\caption{\label{fig:ucu}
$k$-fold uniformly controlled $m$-qubit gate,
$\UC{k}{T}{U(2^m)}$, stands for a sequence of $k$-fold controlled gates $U_i$. Each of the gates acts
on the set of target qubits~$T$. Here $U_i\in U(2^{m})$, where $i = 1, \ldots, 2^k$.}
\end{figure}

Sequences of consequent controlled gates with slightly different control node configurations often appear in quantum circuit diagrams. Let us call a sequence of $2^k$ gates, each having a different sequence of $k$ control nodes, a uniformly controlled $U$ gate, see Fig.~\ref{fig:ucu}. The gate shown acts on the whole $n$-qubit register and, hence, it has $m=n-k$ target qubits denoted by the set $T$. Let us denote a gate of this kind by the symbol $\UC{k}{T}{U(2^m)}$.

The concept of uniformly controlled gates with efficient gate implementation was for the first time introduced in Ref.~\cite{CSD_PRL} in the context of uniformly controlled rotations. It has also been utilized in decompositions of general $n$-qubit gates~\cite{CSD_PRL,pitkapaperi,shende_matrix,tucci2}, and in preparation of quantum states~\cite{stateprep,shende_matrix,pitkapaperi}. Bullock {\it et al.} have generalized uniformly controlled gates for a quantum register which is built of qudits, $d$-level ($d>2$) quantum systems~\cite{bullock_quditl}. The methods to implement uniformly controlled $z$ rotations are also closely related to the earlier work by Bullock and Markov~\cite{bullock:2004}.

Let us construct an elementary gate circuit for a uniformly controlled one-qubit gate. We present the decomposition of uniformly controlled one-parameter rotations, $\UC{k}{t}{R_{\bf a}}$, separately since they require less gates to implement compared with general $\UC{k}{t}{U(2)}$ gates. In a gate $\UC{k}{t}{R_{\bf a}}$ the rotation angles vary, but the rotation axis is the same for each of the subrotations. In the spirit of Eq.~(\ref{eq:diag}), we may assume that the fixed axis ${\bf a}$ is perpendicular to $x$ axis and, hence, we may employ Eq.~(\ref{eq:mopo2}) in the calculations. 

\begin{figure}
\center
\includegraphics[width=0.8\textwidth]{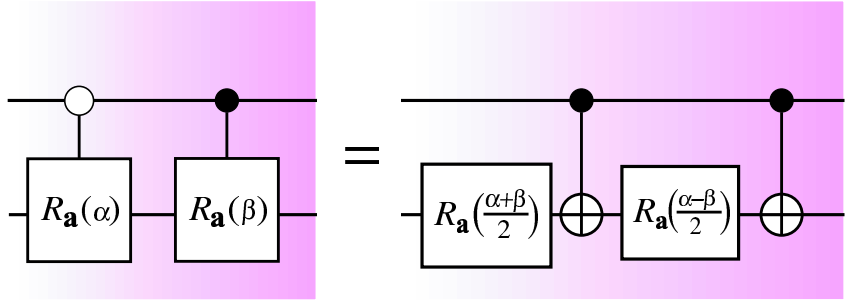}
\caption{\label{elemdec} Implementation of a uniformly controlled one-parameter rotation $R_{\bf a}$ ($a_x=0$) using
the elementary gates.}
\end{figure}

\begin{figure}
\center
\includegraphics[width=0.95\textwidth]{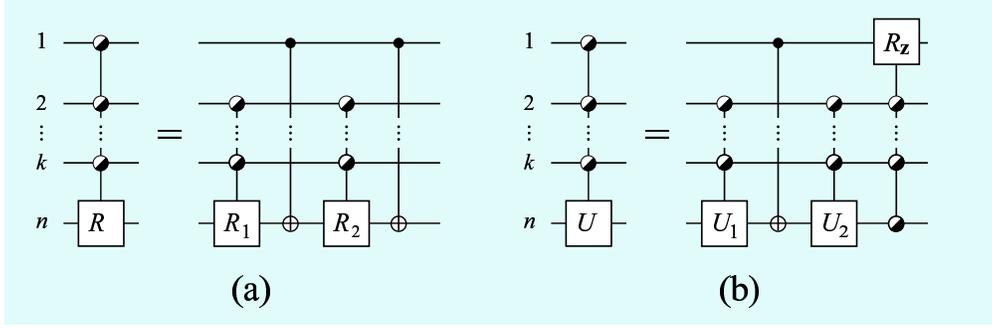}
\caption{\label{fig:rekursioaskel} Decomposition of
a uniformly controlled one-qubit gate. (a) One parameter rotation,
(b) general one-qubit gate $U \in SU(2)$.}
\end{figure}

Figure~\ref{elemdec} shows how to decompose a gate $\UC{1}{2}{R_{\bf a}}$ into two {\footnotesize{CNOT}}s and two elementary rotations. For the states with the the control qubit in state $\ket{0}$ the {\footnotesize{CNOT}}s are inactive and using Eq.~(\ref{eq:mopo}) the rotation angles of the rotations $R_{{\bf a}}(\frac{\alpha+\beta}{2})$ and $R_{{\bf a}}(\frac{\alpha-\beta}{2})$ may be added to obtain the correct rotation $R_{{\bf a}}(\alpha)$. For control qubit sates $\ket{1}$ the rotation $R_{{\bf a}}(\frac{\alpha-\beta}{2})$ is negated according to Eq.~(\ref{eq:mopo2}) and the resulting gate is $R_{{\bf a}}(\beta)$. By adding qubits and control nodes we obtain the general step to eliminate control nodes from the uniformly controlled rotations as shown in Fig.~\ref{fig:rekursioaskel}(a).

\subsection{Decomposition of uniformly controlled one-qubit gates}\label{ucog}

To justify the control node elimination shown in Fig.~\ref{fig:rekursioaskel}(b), we need to introduce so-called constant quantum multiplexor. The idea is that a onefold uniformly controlled rotation is decomposed as \begin{equation} \label{eq:mplex}
\begin{pmatrix}
a \\
  & b
\end{pmatrix}
=
\underbrace{\begin{pmatrix}
r^\dagger \\
  & r
\end{pmatrix}}_{R}
\, \underbrace{\begin{pmatrix}
u \\
  & u
\end{pmatrix}}_{I \otimes u}
\, \underbrace{\begin{pmatrix}
d \\
  & d^\dagger
\end{pmatrix}}_{D}
\, \underbrace{\begin{pmatrix}
v \\
  & v
\end{pmatrix}}_{I \otimes v},
\end{equation}
where $a$, $b$, $u$ and $v$ are unitary and $r$ and $d$ are diagonal unitary $2\times2$ matrices. Here  $a$ and $b$ are fixed by the uniformly controlled gate we are implementing, $u$ and $v$ correspond to the resulting one-qubit gates and the uniformly controlled $z$ rotation corresponding to matrix $r$ is to be tuned such that the diagonal matrix $d$ separating the one qubit gates is independent of $a$ and $b$.

Equation~(\ref{eq:mplex}) yields the matrix equations
\begin{align}
a &= r^\dagger u d v, \\
b &= r u d^\dagger v
\end{align}
or, equivalently,
\begin{align}
\label{eq:ab}
X &:= a b^\dagger = r^\dagger u d^2 u^\dagger r^\dagger, \\
v &= d u^\dagger r^\dagger b = d^\dagger u^\dagger r a. \label{eq:v}
\end{align}
Equation~(\ref{eq:ab}) may be recast into a form reminiscent of an eigenvalue
decomposition:
\begin{equation}
r X r = u d^2 u^\dagger =:  u \Lambda u^\dagger.
\end{equation}
Note that $X$ is fixed by the matrices $a$ and $b$, but $r$ can be chosen freely. By diagonalizing the matrix $r X r$, we find the similarity transformation~$u$ and the eigenvalue matrix $\Lambda = d^2$. The matrix $v$ is obtained by inserting the results into Eq.~(\ref{eq:v}).

\begin{figure}\begin{center}
\includegraphics[width=0.55\textwidth]{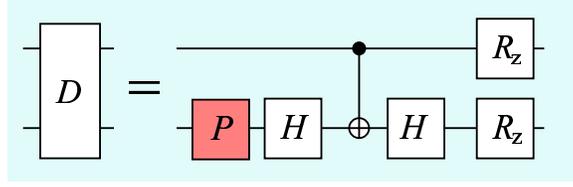}
\caption{\label{fig:d} Elementary gate sequence for the $D$ gate, where
$H$ is the Hadamard gate and $R_z=R_z(\pi/2)$. Gate $P=e^{-i \pi/4}$ is an
adjustment of the global phase.}\end{center}
\end{figure}

Since $X \in U(2)$, we may express it using the parametrization
\begin{equation}
X =
\begin{pmatrix}
x_1 & x_2 \\
-\bar{x}_2 & \bar{x}_1
\end{pmatrix}
e^{i \phi / 2},
\end{equation}
where $|x_1|^2 + |x_2|^2 = 1$ and $\det(X) = e^{i \phi}$.
The characteristic polynomial of the matrix $r X r$ is
\begin{equation}
\det(r X r - \lambda I) = \lambda^2 - \lambda \left(r_1^2 x_1 + r_2^2 \bar{x}_1 \right) e^{i \phi /
2} + r_1^2 r_2^2 e^{i \phi}.
\end{equation}
Let us fix the freely tunable matrix $r$ to be
\begin{equation}
r=
\begin{pmatrix}
e^{\frac{i}{2} \left[ -\frac{\pi}{2} - \frac{\phi}{2} - \arg(x_1) \right]} & \\
& e^{\frac{i}{2} \left[  \frac{\pi}{2} - \frac{\phi}{2} + \arg(x_1) \right]}
\end{pmatrix},
\end{equation}
which implies the matrix $d$ to be, indeed, independent of the matrices $a$ and $b$. Namely
\begin{equation}
\Lambda = d^2 =
\begin{pmatrix}
e^{i \frac{\pi}{2}} & \\
& -e^{i \frac{\pi}{2}}
\end{pmatrix}.
\end{equation}
Hence, the diagonal multiplexing gate~$D$ obtains the fixed form $D~=~e^{i \frac{\pi}{4} \sigma_z \otimes \sigma_z}$, which can be realized straightforwardly using an Ising-type Hamiltonian or, alternatively, it can be decomposed into a {\footnotesize{CNOT}} and one-qubit gates as shown in Fig.~\ref{fig:d}.

\begin{figure}\begin{center}
\includegraphics[width=0.55\textwidth]{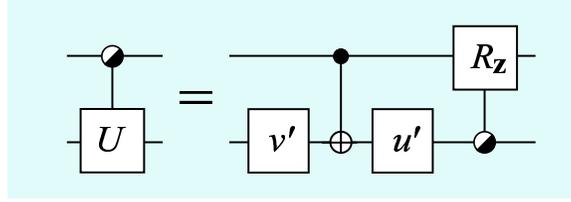}
\caption{\label{fig:qm_rz} Constant quantum multiplexor for two
qubits.}\end{center}
\end{figure}

The single qubit gates acting on the bottom qubit in Fig.~\ref{fig:d} may be merged with the adjacent single qubit gates $u$ and $v$ resulting in single qubit gates $u'$ and $v'$ shown in Fig.~\ref{fig:qm_rz}, respectively. The $z$ rotation acting on the top qubit in Fig.~\ref{fig:d} may be correspondingly merged with the uniformly controlled $z$ rotation in Fig.~\ref{fig:qm_rz} and, hence, we have justified elimination of the control node for a onefold uniformly controlled one-qubit gate shown in Fig.~\ref{fig:qm_rz}. By adding qubits with control nodes we obtain the general step to eliminate control nodes from the UCGs as shown in Fig.~\ref{fig:rekursioaskel}(b).

\begin{figure}
\center
\includegraphics[width=0.95\textwidth]{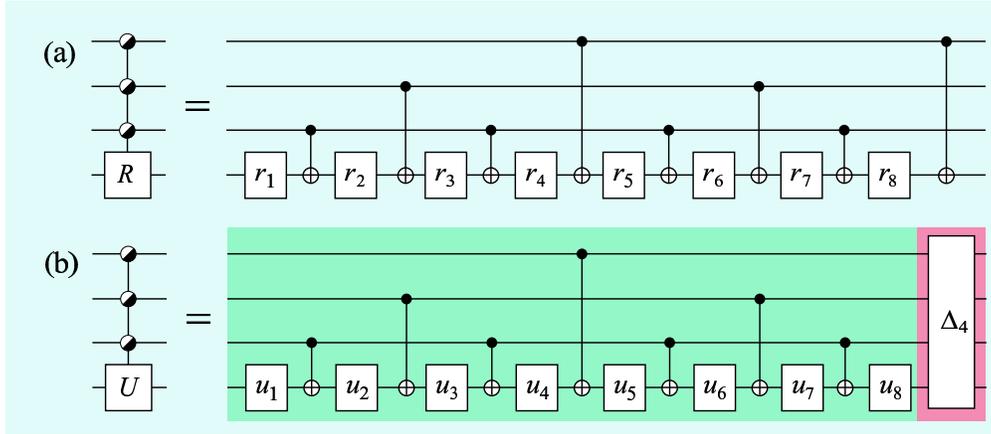}
\caption{\label{fig:3tasaistaaskelta} Quantum circuit realizing a threefold uniformly controlled
(a) one-parameter rotation, (b) general one-qubit gate.
In (a) $\{r_i\}$ stand for a one-parameter rotations and in (b) $\{u_i\}$ are general one-qubit gates.
Here the gate $\Delta_4$ corresponds to a diagonal $16 \times 16$ unitary  matrix.}
\end{figure}

We note that the uniformly controlled rotations in Fig.~\ref{fig:rekursioaskel}(a) have the same rotation axis and, hence, commute. Thus the first uniformly controlled rotation may be, as well, transfered to be the last gate. We call this procedure mirroring the circuit. By using the step of Fig.~\ref{fig:rekursioaskel}(a) recursively and mirroring every second outcome of the recursion we obtain the full decomposition of $\UC{k}{t}{R_{\bf a}}$ using only $2^k$ one-qubit rotations $R_{\bf a}$ and the same number of {\footnotesize{CNOT}}s. An example of the case $k=3$ is shown in Fig.~\ref{fig:3tasaistaaskelta}(a). When decomposing general one-qubit UCGs, the step in Fig.~\ref{fig:rekursioaskel}(b) is to be used recursively. There we have to keep in mind that, actually, the {\footnotesize{CNOT}} may be taken to be the diagonal gate $D$ show in Fig.~\ref{fig:d}. Hence, when the recursion is applied always on the leftmost UCG, all the resulting uniformly controlled $z$ rotations may be merged with the adjacent UCGs except the rightmost ones which pile on to form a diagonal matrix $\Delta_4$. The decomposition of $\UC{3}{4}{U(2)}$ is shown in Fig.~\ref{fig:3tasaistaaskelta}(b).

In general, the decomposition of a gate $\UCU{k}{t}$ includes an alternating sequence of $2^k$~one-qubit gates and $2^k-1$~{\footnotesize{CNOT}}s which we denote by $\TUCU{k}{t}$. Likewise, the implementation involves a cascade of $k$~distinct uniformly controlled $z$~rotations which corresponds to a single diagonal $(k+1)$-qubit gate $\Delta_{k+1}$. However, the implementation of this part of the gate sequence can often be circumvented by merging it with the adjacent gates as shown in Sec.~\ref{csd}. In fact, if the qubit register is measured as such after the action of the gate $\UCU{k}{t}$, the diagonal gate may be left unimplemented since it does not change the probability amplitudes.

\subsection{Nearest-neighbor decompositions}\label{ucnn}

In the practical realization of a quantum computer, the spatial arrangement of qubits or other reasons may limit the interactions between the qubits. Let us consider a quantum register whose topology corresponds to that of a linear chain and which allows the gates to act only on nearest-neighbor qubits. This topology turns out to be amenable for implementing a uniformly controlled gate, which may have important consequences for experimentally realizing quantum computing.

\begin{figure}
\center
\includegraphics[width=0.45\textwidth]{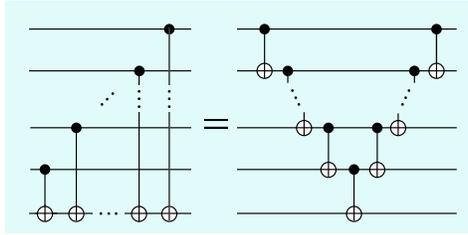}
\caption{\label{fig:kaskaadi} {\footnotesize{CNOT}} cascade which can be efficiently
implemented using nearest-neighbor {\footnotesize{CNOT}}s~\cite{tucci3}.}
\end{figure}

\begin{figure}
\center
\includegraphics[width=0.95\textwidth]{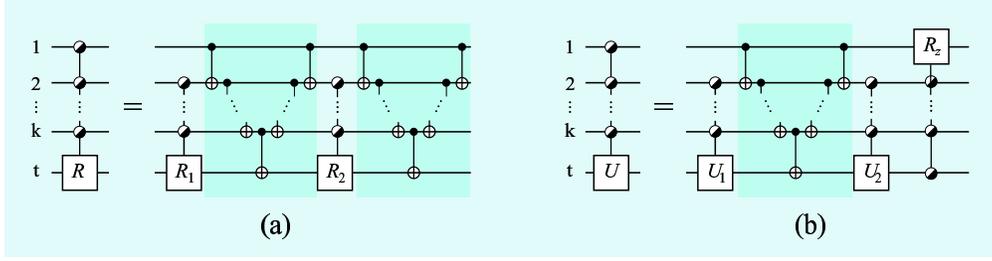}
\caption{\label{fig:rzaskel} Method for reducing a uniformly controlled gate into nearest-neighbor gates:
(a) uniformly controlled rotation and
(b) general one-qubit gate.
Here the circuit diagrams may also be mirrored horizontally.
}
\end{figure}

The quantum circuit presented for a uniformly controlled gate can be translated efficiently into an array of nearest-neighbor gates. The technique is based on the circuit identity shown in Fig.~\ref{fig:kaskaadi}.  The strategy is to modify the decomposition shown in Fig.~\ref{fig:rekursioaskel} by inserting an identity in the form of a {\footnotesize{CNOT}} cascade and its inverse, a similar cascade, into the circuit next to each {\footnotesize{CNOT}}. The inverse cascades are absorbed into the adjacent uniformly controlled gate. The remaining cascades, together with the original {\footnotesize{CNOT}}s, can be efficiently implemented using nearest-neighbor {\footnotesize{CNOT}}s as illustrated in Fig.~\ref{fig:kaskaadi}. The control node elimination steps for the nearest-neighbor implementation are shown in Fig.~\ref{fig:rzaskel}.

The complexity of the nearest-neighbor implementation depends on the relative order of the target and control qubits, and the order in which the control qubits are eliminated. An efficient strategy is to first eliminate the control nodes that are furthest apart from the target. Furthermore, for the gates with numerous control nodes, it is advantageous to use a sequence of swap gates to move the target qubit next to the center of the chain before the operation and back after it. A swap gate can be realized using three consecutive {\footnotesize{CNOT}}s~\cite{NielsenChuang}.

\begin{figure}
\center
\includegraphics[width=0.95\textwidth]{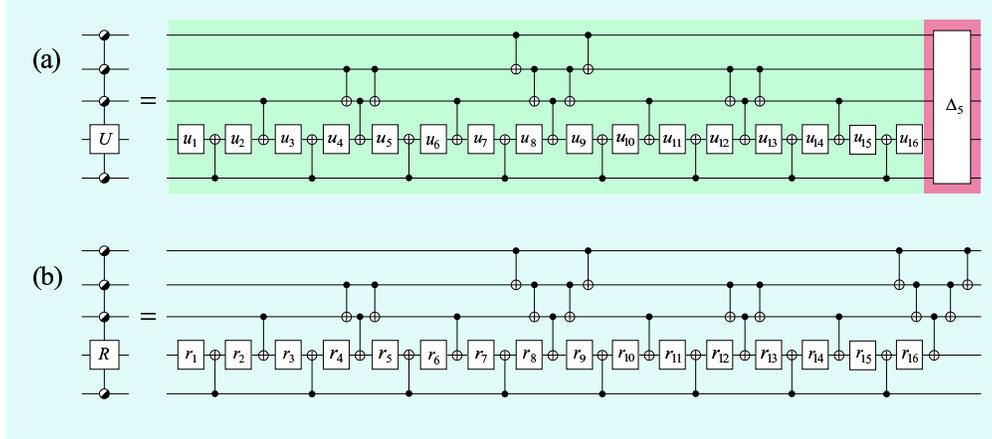}
\caption{\label{fig:tasainenlahinaapuri} Implementation of threefold uniformly controlled (a) general one-qubit gate and (b) one-parameter rotation. Here gates $\{r_i\}$ are generic rotations about $x$ axis and gates $\{u_i\}$ belong to $SU(2)$. The alternating sequence of {\footnotesize{CNOT}}s and $u_i$ gates is denoted by $\TUCU{4}{5}$. The rightmost sequence of uniformly controlled $z$ rotations corresponds to a single diagonal gate, denoted by $\Delta_5$.
}
\end{figure}

Using this strategy a gate $\TUCU{n-1}{t}$ can be implemented with at most
\begin{equation} \label{eq:ucu-nnCNOTs}
C_{U(2)}(n, s) = \frac{5}{6} 2^{n} +2n-6s
- \left\{\begin{matrix}
\frac{1}{3}, \quad \text{$n$ even} \\
\frac{5}{3}, \quad \text{$n$ odd}
\end{matrix} \right.
\end{equation}
nearest-neighbor {\footnotesize{CNOT}}s. Here $s = 1, \ldots, \lceil\frac{n}{2} \rceil$ is the distance of the target qubit~$t$ from the end of the chain. Figure~\ref{fig:tasainenlahinaapuri}(a) depicts the resulting circuit for the case $k=4$ and $s=1$. Similar treatment for gate $\UC{n-1}{t}{R_{\bf a}}$ yields a quantum gate array with
\begin{equation} \label{eq:ucr-nnCNOTs}
C_{R}(n, s) = \frac{5}{6} 2^{n} +3n-6s
-\left\{\begin{matrix}
\frac{4}{3}, \quad \text{$n$ even} \\
\frac{5}{3}, \quad \text{$n$ odd}
\end{matrix} \right.
\end{equation}
nearest-neighbor {\footnotesize{CNOT}}s. Figure~\ref{fig:tasainenlahinaapuri}(b) displays an example circuit for the case $k=4$ and $s=1$.

We note that the uniformly controlled one-qubit gate carries $3\cdot 2^k$ degrees of freedom, and requires roughly the same number of elementary gates for its implementation. Thus an array of nearest-neighbor {\footnotesize{CNOT}}s provides an efficient implementation for uniformly controlled one-qubit gates, and therefore for any uniformly controlled gate. In particular this can be utilized to efficiently implement unstructured unitary transformations. Furthermore, the structure of the nearest-neighbor circuit allows several gate operations to be executed in parallel which may further reduce the execution time of the algorithm.

\section{QR decomposition}\label{qr}

Numerical matrix computation~\cite{golub} is a field of mathematics that provides useful tools to construct and manipulate quantum gate arrays. For example, the theorem of QR decomposition states that for each complex matrix $A$ there exists a unitary matrix $Q$ and an upper triangular matrix $R$ such that $A=QR$. Here the matrix $Q$ may be a product two-level matrices called Givens rotations~\cite{Givens}. For unitary matrices $A$, the resulting matrix $R$ is essentially an identity. Thus the sequence of Givens rotations yields a decomposition of any unitary matrix into two-level matrices. Consequently, these two-level matrices may be decomposed into elementary gates as shown below. Traditionally, a technique based on this principle is employed in quantum computation to find the elementary decomposition of an unstructured unitary matrix~\cite{barenco,cybenko,aho}. Reference~\cite{QR_PRL} presents improvements to the traditional construction that eventually lead to the quantum gate decomposition of minimal complexity $O(4^n)$.

Let us outline how to find the sequence of Givens rotations, the  product of whom implements any unitary matrix $U \in SU(2^n)$. In the case $n=1$, a Givens rotation $G\in SU(2)$ corresponding to a vector ${\bf b}=(b_1\, b_2)^T$ may be defined as
\beq\label{giev}
G\,{\bf b}=\frac{1}{\sqrt{|b_1|^2+|b_2|^2}}
\begin{pmatrix}
b_1^* & b_2^* \\
-b_2 & b_1
\end{pmatrix}
\begin{pmatrix}
b_1 \\
b_2
\end{pmatrix}
=
\begin{pmatrix}
\sqrt{|b_1|^2+|b_2|^2} \\
0
\end{pmatrix}.
\eeq
For general number of qubits $n$, a Givens rotation is a two-level matrix acting non-trivially only on a subspace spanned by two basis vectors, for example, $\ket{e_j}$ and $\ket{e_k}$. When a Givens rotation is used to nullify elements of a matrix $U\in SU(N)$ we also need to specify the column which is used as the vector corresponding to the rotation. Hence, we define a Givens rotation $^iG_{j,k}$ to be a two-level complex matrix which selectively nullifies the element on the $i^{\rm th}$ column and the $j^{\rm th}$ row of the matrix $U$ against the element on the $i^{\rm th}$ column and the $k^{\rm th}$ row. For example
\beq
{}^1G_{N,N-1} U= \!\left(%
\begin{array}{cccc}
u_{1,1} & u_{1,2} & \ldots & u_{1,N}  \\
\vdots & \vdots & \ddots &\vdots  \\
u_{N-2,1} & u_{N-2,2} & \ldots & u_{N-2,N}  \\
\tilde{u}_{N-1,1} & \tilde{u}_{N-1,2} & \ldots & \tilde{u}_{N-1,N}  \\
0 & \tilde{u}_{N,2} & \ldots & \tilde{u}_{N,N}  \\
\end{array}%
\right),
\eeq
where the elements of $\tilde{U}$ that differ from those of $U$ are indicated with the tilde.

Applying ${}^1G_{N-1,N-2}$ to the modified matrix $\tilde{U}$ we can nullify the element $\tilde{u}_{N-1,1}$ and similarly the whole first column, except the diagonal element. The unitarity of the matrix $U$ fixes its absolute value to unity and the definition of a Givens rotation in Eq.~(\ref{giev}) assures that the phase of the diagonal element vanishes, {\it i.e.}, it obtains value 1. The further application of the method to the columns from $2$ to $N-1$ results in an identity matrix as
\beq\label{eq:gu}
\left(\prod_{i=1}^{2^n-1}\prod_{j=i+1}^{2^n}
{^{2^n-i}}G_{j,j-1}\right)U= I,
\eeq
where the product of the non-commuting matrices is taken from left to right as always in this chapter. Equation~(\ref{eq:gu}) yields the factorization of the arbitrary matrix $U\in SU(2^n)$ using Givens rotations
\beq
U=
\left(\prod_{i=1}^{2^n-1}\,\prod_{j=1}^{2^n-i}
{^{i}}G_{2^n-j+1,2^n-j}^{\dagger}\right),
\eeq
which introduces an implementation of an arbitrary quantum gate provided that an elementary gate presentation of each of the Givens rotations is known. We note the non-zero off-diagonal elements of ${^i}G_{j,k}$ by $2\times 2$-matrix ${^i}\Gamma_{j,k}$.

In the first presentation of the QR decomposition for arbitrary quantum gates~\cite{barenco}, the basis vectors were ordered using standard binary coding. Thus the Givens rotations acting on adjacent basis vectors do not directly correspond to any known gate. However, if the basis vectors are permuted before the action of every rotation and permuted back after the action, the rotations may be written as fully controlled one-qubit gates. The permutation for each $O(4^n)$ rotations needed of the order of $n$ fully controlled~{\footnotesize{NOT}} gates each of which required of the order of $n^2$ {\footnotesize{CNOT}}s. Hence, the complexity of the whole decomposition turned out to be $O(n^34^n)$.

\begin{figure}
\center
\includegraphics[width=0.75\textwidth]{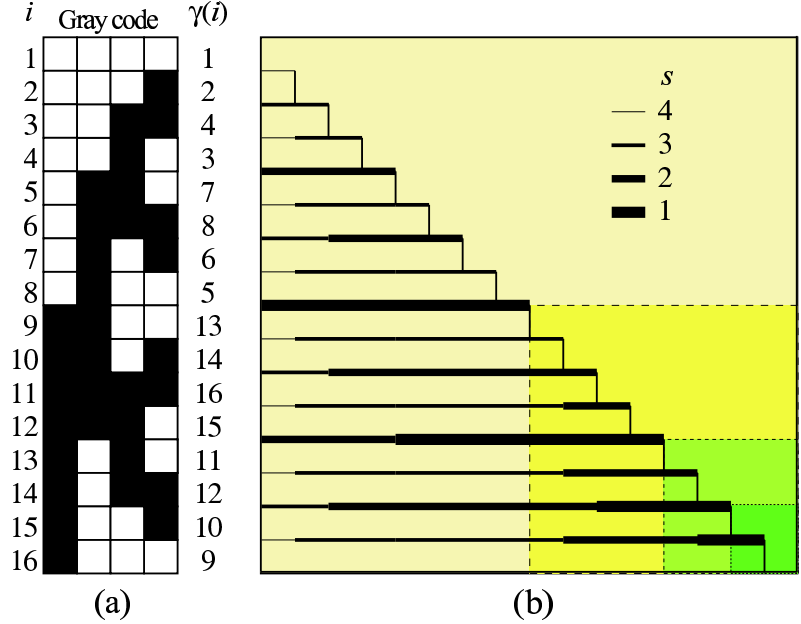}
\caption{\label{fg:peli} (a) Four bit Gray code.  White squares stand for bit values $0$ and black squares denote $1$. (b) The number of control nodes needed for the Givens rotation nullifying the elements of the matrix $U$. The width of the line $s$ between the matrix elements represents the number of control nodes which may be eliminated.
}
\end{figure}

Instead of labelling the basis vectors using standard binary coding, the binary reflected Gray code was employed in Ref.~\cite{QR_PRL}. The special property of any Gray code ordered basis is that only one bit changes between the adjacent basis vectors $\ket{e_i}$ and $\ket{e_{i+1}}$, see Fig.~\ref{fg:peli}(a). The important consequence of this is that the operations limited to the subspace spanned by $\ket{e_i}$ and $\ket{e_{i+1}}$ take the form of a C$^{n-1}V$ gate, where $V \in SU(2)$. Consequently, each of the $2^{n-1}(2^n-1)$ Givens rotations $\{{^i}G_{j,j-1}\}$ can be implemented using only one C$^{n-1}V$ gate and no basis permuting gates are needed between them. Since a C$^{n-1}V$ gate may be decomposed into $O(n)$ {\footnotesize{CNOT}}s~\cite{barenco}, the decomposition has a complexity $O(n4^n)$ at this point. We note that actually we may, as well, label the basis vectors using the standard binary coding but, instead, the order in which the elements of the matrix $U\in SU(2^n)$ are nullified must be chosen such that the Givens rotations operate non-trivially only to basis vectors with binary presentations differing only in one bit. Provided that the basis vectors are labelled using the standard binary coding, the matrices ${^{2^n-i}}G_{j,j-1}$ in Eq.~(\ref{eq:gu}) become ${^{\gamma(2^n-i)}}G_{\gamma(j),\gamma(j-1)}$, where the function $\gamma(i)$ gives the integer value of the $i^{th}$ element in the binary reflected Gray code, see Fig.~\ref{fg:peli}.

Furthermore, we find that only a small fraction of the control nodes in the fully controlled one-qubit gates appears to be essential for the final result of the decomposition. If $s$ control nodes are removed from a C$^{n-1}({^i}\Gamma_{j,j-1})$ gate, the matrix representation $^iG_{j}^s$ of such an operation is no more two-level, but rather $2^{s+1}$-level, {\it i.e.}, the matrix $^iG_{j}^s$ operates with the matrix $^i\Gamma_{j,j-1}$ to all pairs of basis vectors which satisfy the remaining control conditions. Once some element of the matrix $U$ we are decomposing becomes zero in the diagonalization process, we must remove control nodes from the following fully controlled gates in such a way that the zeroed element does not mix with the non-zero elements.

\begin{figure}
\center
\includegraphics[width=1.00\textwidth]{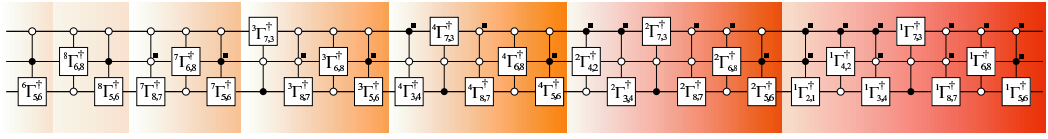}
\caption{\label{fg:3qubitfull} Quantum circuit equivalent to an
arbitrary three-qubit quantum gate $U\in SU(8)$. The
control nodes indicated with a black square on the upper right hand
side corner are superfluous and may be omitted to decrease the
complexity of the decomposition.}
\end{figure}

Figure~\ref{fg:peli}(b) illustrates the determination of the number of control nodes necessary in the diagonalization. The total number of gates in the implementation depends on the number of the control nodes in each of the involved gates. Let us denote by $g_n(k)$ the number of C$^kV$ gates requires for the whole diagonalization process of an $n$-qubit gate. In Ref.~\cite{QR_PRL}, a recursion relation for $g_n(k)$ was derived. The relation has an awkward analytic solution and, therefore, it was estimated from above as
\beq\label{app}
g_n(n-i)\leq 2^{n+i}.
\eeq
Equation~(\ref{app}) shows that the number of $k$-fold controlled gates decreases exponentially with the number of control nodes. On the other hand, gate C$^kV$ takes $O(n)$ gates to implement~\cite{barenco}. These results together imply that the gate array for an $n$-qubit unitary gate involves $O(4^n)$ elementary gates. Figure~\ref{fg:3qubitfull} shows an example of the quantum circuit equivalent to an arbitrary three-qubit quantum gate $U\in SU(8)$.

To calculate the number of elementary gates, we use the decompositions described in Ref.~\cite{barenco}. For large $n$, the leading contribution to the number of {\footnotesize{CNOT}}s is approximately $8.7\times 4^n$, while the upper bound from Eq.~(\ref{app}) yields approximately $11\times 4^n$. We note that neither one of the two techniques alone, the Gray code ordered basis vectors nor the elimination of the control nodes suffices to decrease the circuit complexity to $O(4^n)$. As a curiosity, the technique to eliminate control nodes has recently been generalized and adopted again to the numerical matrix computation~\cite{oleary}.

\section{Cosine-sine decomposition}\label{csd}
\subsection{Recursive cosine-sine decomposition}\label{rcsd}
The CSD of a unitary $2^n \times 2^n$ matrix may be expressed as~\cite{Paige}
\begin{equation}\label{eq:cs}
U =
\underbrace{\begin{pmatrix}
u_{1} & 0\\
0 & u_{2}
\end{pmatrix}}_{U_1}
\underbrace{
\begin{pmatrix}
c & s \\
-s & c
\end{pmatrix}
}_{A}
\underbrace{\begin{pmatrix}
u_{3} & 0\\
0& u_{4}
\end{pmatrix}}_{{U}_2},
\end{equation}
where $\{u_{k}\}$ are unitary $2^{n-1} \times 2^{n-1}$ matrices and the real diagonal matrices $c$ and $s$ are of the form $c = \text{diag}_{l}(\cos \theta_{l})$ and $s = \text{diag}_{l}(\sin \theta_{l})$ ($l = 1, \ldots, 2^{n-1}$). The matrix $A$ corresponds to a uniformly controlled $y$ rotation $\UC{n-1}{1}{R_y}$ with rotation angles $\{\theta_l\}$ and the matrices $U_1$ and $U_2$ to uniformly controlled $(n-1)$-qubit gates $\UC{1}{T}{SU(2^{n-1})}$. By applying Eq.~(\ref{eq:cs}) recursively to the uniformly controlled multi-qubit gates until we only have uniformly controlled one-qubit gates, we obtain a decomposition
\begin{equation}
\label{eq:csdres}
U(2^n) = \UCU{n-1}{n} \prod_{i=1}^{2^{n-1}-1} \UC{n-1}{n-\zeta(i)}{R_y} \UCU{n-1}{n},
\end{equation}
where $\zeta$ is the ruler function~\cite{ruler}.

\begin{figure}
\center
\includegraphics[width=0.90\textwidth]{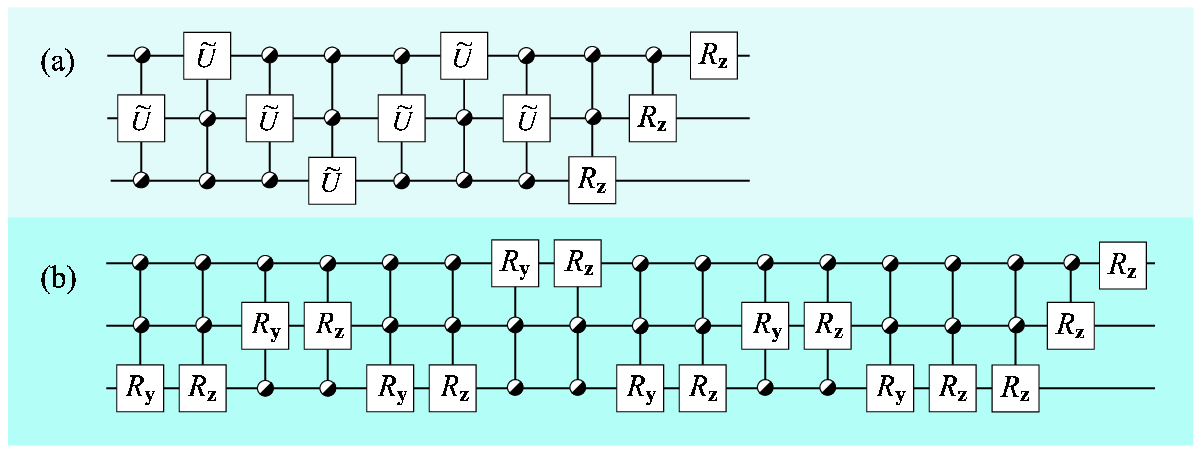}
\caption{\label{fg:kolmenpiirit} Quantum circuit for a three-qubit gate obtained using (a) the current CSD and (b) an alternative CSD.}
\end{figure}

We begin to decompose the rightmost gate in Eq.~(\ref{eq:csdres}) into elementary gates by writing an identity $I=\Delta_n\Delta_n^*$ between the gates $\UC{n-1}{n-\zeta(2^{n-1}-1)}{R_y}$ and $\UCU{n-1}{n}$. Here we choose $\Delta$ such that 
\begin{equation}
\UCU{n-1}{n} = \Delta_n \TUCU{n-1}{n},
\end{equation}
where the gate $\TUCU{n-1}{n}$ introduced in Sec.~\ref{ucog} needs only $2^{n-1}-1$ {\footnotesize{CNOT}}s to implement. We are now left with the product
\begin{eqnarray}
\label{eq:csdres}
U(2^n) &=& \UCU{n-1}{n} \left[\prod_{i=1}^{2^{n-1}-2} \UC{n-1}{n-\zeta(i)}{R_y} \UCU{n-1}{n}\right] \\
 &&\times\UC{n-1}{n-\zeta(2^{n-1}-1))}{R_y}\Delta_n \TUCU{n-1}{n},
\end{eqnarray}
where the product $\UC{n-1}{n-\zeta(2^{n-1}-1))}{R_y}\Delta_n$ may be written as a single uniformly controlled one-qubit gate $\UCU{n-1}{n-\zeta(2^{n-1}-1))}$. Continuing to change the $\UCU{n-1}{}$ gates into $\TUCU{n-1}{}$ gates by adding diagonal gates, we finally obtain the decomposition
\begin{equation}
\label{eq:csdres2}
U(2^n) = \Delta_n \TUCU{n-1}{n} \prod_{i=1}^{2^{n-1}-1}
\TUCU{n-1}{n-\gamma(i)} \TUCU{n-1}{n}.
\end{equation}
There are $2^n-1$ $\TUCU{n-1}{}$ gates in Eq.~(\ref{eq:csdres2}), each of which may be decomposed into $2^{n-1}-1$ {\footnotesize{CNOT}}s. In addition, we have to implement the diagonal gate $\Delta_n$ using $2^n-2$ {\footnotesize{CNOT}}s~\cite{bullock:2004}. Actually, one more {\footnotesize{CNOT}} may be eliminated~\cite{pitkapaperi} and, hence, the current CSD requires $4^n/2-2^{n-1}-2$ {\footnotesize{CNOT}}s. The circuit diagram for the CSD in the case $n=3$ is shown in Fig.~\ref{fg:kolmenpiirit}(a), where the diagonal gate $\Delta_3$ is written as a cascade of uniformly controlled $z$ rotations~\cite{CSD_PRL}. There exists also a slightly different version of the CSD where the matrix $U\in SU(2^n)$ is decomposed only into uniformly controlled $z$ and $y$ rotations~\cite{CSD_PRL}. An example of this alternative CSD in shown in Fig.~\ref{fg:kolmenpiirit}(b). Actually, the alternative CSD is obtained also from the current one by writing the rightmost UCG in the product of Eq.~(\ref{eq:csdres}) as a product $\UC{n-1}{}{R_z}\UC{n-1}{}{R_y}\UC{n-1}{}{R_z}$. Being diagonal, the gate $\UC{n-1}{}{R_z}$ may be merged into the adjacent UCG and the process can be continued until the last $\UC{n-1}{}{R_z}$ arising from the leftmost UCG may be merged to the diagonal gate $\Delta_n$.

\subsection{Top down approach}
\begin{figure}
\center
\includegraphics[width=0.95\textwidth]{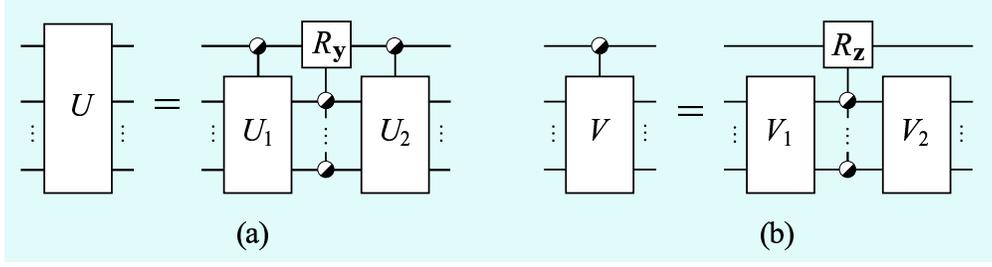}
\caption{\label{fg:csdqm} Circuit diagram for (a) cosine-sine decomposition and (b) Quantum multiplexor.}
\end{figure}

In addition to the CSD described above, Ref.~\cite{shende_matrix} presents an alternative approach employing the cosine-sine decomposition. This method is called NQ decomposition\footnote{NQ stands for $n$ qubits.} and it is almost as efficient as the CSD discussed in Sec.~\ref{rcsd}. The first step of the NQ method is the same as in the CSD shown in Fig.~\ref{fg:csdqm}(a), see also Eq.~(\ref{eq:cs}). However, the CSD step is not used recursively but, instead, the control nodes in the UCG are eliminated using quantum multiplexor shown in Fig.~\ref{fg:csdqm}(b). After the application of the CSD and the quantum multiplexor, we are left with three uniformly controlled rotations separating four uncontrolled $n-1$ qubit gates. Since the NQ step produces pure gates acting on fewer qubits it is also called a top down approach.

Let us now motivate the validity of the quantum multiplexor. It is very similar to the constant quantum multiplexor presented in Sec.~\ref{ucog} but the matrices corresponding to Eq.~(\ref{eq:mplex}) are $2^{n-1}\times 2^{n-1}$-dimensional, the matrix $r$ is omitted and the diagonal matrix $d$ may depend and the matrices $a$ and $b$. Actually, it is an open problem whether there exists a constant quantum multiplexor in this general case, {\it i.e.}, can we find diagonal $r\in SU(2^n)$ for every $X\in SU(2^n)$ such that the eigenvalues of $rXr$ are fixed. The matrix equation corresponding to Fig.~\ref{fg:csdqm}(b) reads as
\beq\label{mmpl}
\begin{pmatrix}
a \\
  & b
\end{pmatrix}
=
\, {\begin{pmatrix}
u \\
  & u
\end{pmatrix}}
\, {\begin{pmatrix}
d \\
  & d^\dagger
\end{pmatrix}}
\, {\begin{pmatrix}
v \\
  & v
\end{pmatrix}},
\eeq
where $a$, $b$, $u$ and $v$ are unitary and $d$ is diagonal
unitary $2^{n-1}\times2^{n-1}$ matrices. 
Equation~(\ref{mmpl}) yields the matrix equations
\begin{align}
a &= u d v, \\
b &= u d^\dagger v
\end{align}
or, equivalently,
\begin{align}
\label{eq:mab}
a b^\dagger &= u d^2 u^\dagger, \\
v &= d u^\dagger b = d^\dagger u^\dagger a. \label{eq:mv}
\end{align}
By diagonalizing the matrix $a b^\dagger$, we find the similarity
transformation~$u$ and the eigenvalue matrix $d^2$.
The matrix $v$ is obtained by inserting the results into Eq.~(\ref{eq:mv}). Hence, we have proven the quantum multiplexor in Fig.~\ref{fg:csdqm}(b).

\begin{figure}
\center
\includegraphics[width=0.92\textwidth]{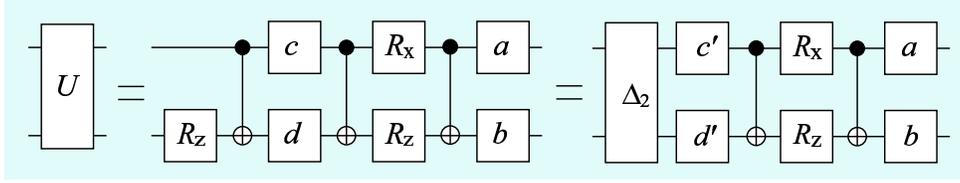}
\caption{\label{fg:u2} The minimal elementary gate construction for a two-qubit gate~\cite{shende}.}
\end{figure}

The NQ step is continued recursively to all the gates except the uniformly controlled rotations until the two-qubit level is encountered. The two-qubit gates are decomposed using the minimal elementary gate construction shown in Fig.~\ref{fg:u2}. In fact, diagonal gates commute with the control nodes of the UCG and, hence, all but one of the resulting two-qubit gates may be implemented up to diagonal, {\it i.e.}, using only two {\footnotesize{CNOT}}s as shown in the leftmost part of Fig.~\ref{fg:u2}. 

We will now calculate the number of the {\footnotesize{CNOT}}s involved in to NQ decomposition of an unstructured $U\in SU(2^n)$. Let us denote this number by $a_n$. Since the NQ step produces four unstructured gates in $SU(2^{n-1})$ and three $\UC{n-1}{}{R}$ gates each of which may be implemented using $2^{n-1}$ {\footnotesize{CNOT}}s, we obtain a recursion relation
\beq
a_n = 4a_{n-1}+\frac{3}{2}2^n.
\eeq 
Using the above discussed condition $a_2=2$ and adding one {\footnotesize{CNOT}} from the only two-qubit gate which needs three {\footnotesize{CNOT}}s for its implementation we obtain the result
\beq
a_n = \frac{1}{2}4^n-\frac{3}{2}2^n+1.
\eeq
Thus compared with the number of {\footnotesize{CNOT}}s from the CSD $\frac{1}{2}4^n-\frac{1}{2}2^n-2$, the NQ decomposition yields the same result in the leading order. However, when we compare the number on one-qubit gates or alternatively elementary rotations, the CSD is found to be more efficient, see Table.~\ref{t:comp}.

\begin{table}[h]
\center
\begin{tabular}{l|c|c}
Gate type & NQ & CSD \\
\hline
{\footnotesize{CNOT}} & $\frac{1}{2}4^n -\frac{3}{2}2^n + 1$ &
$\frac{1}{2}4^n -\frac{1}{2}2^n - 2$ \\
$R_y, R_z$ & $\frac{9}{8}4^n -\frac{3}{2}2^n + 3$ & $4^n - 1$ \\
or $SU(2)$ & $\frac{17}{24}4^n -\frac{3}{2}2^n - \frac{1}{3}$ &
$\frac{1}{2}4^n +\frac{1}{2}2^n - n - 1$
\end{tabular}
\caption{\label{t:comp} Comparison of the gate counts required to
implement a general $n$-qubit gate using the NQ
decomposition~\cite{shende_matrix} and the recursive CSD for unstructured $n$-qubit gates.
}
\end{table}

Actually, the number of {\footnotesize{CNOT}}s in the NQ decomposition may be reduced by noting that the resulting uniformly controlled $y$ rotations may be always implemented up to a diagonal gate as seen from Fig.~\ref{fg:csdqm}(a). Since $\TUCU{k}{}$ gate needs one {\footnotesize{CNOT}} less to implement than $\UC{k}{}{R_y}$ we obtain a recursion relation
\beq
a_n = 4a_{n-1}+\frac{3}{2}2^n-1,
\eeq 
the solution of which is found to be
\beq
a_n = \frac{23}{48}4^n-\frac{3}{2}2^n+\frac{1}{3}
\eeq 
This result is the first known to require less than $\frac{1}{2}4^n$ {\footnotesize{CNOT}}s in the leading order.

\section{Local state preparation}\label{lsp}

The execution of any quantum algorithm requires a certain initial state as an input. Depending on the physical realization of the quantum computer, convenient initialization procedures may only produce a limited range of states possibly not containing the desired initial state. This brings up the problem of local state preparation\footnote{We use the word local to separate the state preparation discussed here from the remote state preparation related to quantum teleportation.}, {\it i.e.}, how to implement the transformation of an arbitrary quantum state into another one.

The configuration space of the $n$-qubit quantum register is $2^n$-dimensional complex space. Excluding the global phase and state normalization, we find that the general unitary transformation transforming a given $n$-qubit state into another must have at least $2 \times 2^n - 2$ real degrees of freedom. Hence, in the worst-case scenario, the corresponding quantum circuit should involve at least $2^{n+1} - 2$ elementary rotations, each carrying one degree of freedom. Since each of the {\footnotesize{CNOT}}s can bind at most four elementary rotations~\cite{shende}, at least $\lceil \frac{1}{4} (2^{n+1} - 3n - 2) \rceil$ of them are needed. However, no quantum circuit construction embodying the minimal complexity has been presented in the literature. An upper bound for the number of gates needed for state preparation has been considered by Knill~\cite{knill}, who found that no more than $O(n2^n)$ gates provide the circuit implementing the transformation. More recently, a sufficient circuit of $O(2^n)$ elementary gates was obtained in Ref.~\cite{Shende_vector} (see also Ref.~\cite{stateprep}) as a special case of the method developed for QR decomposition of a general quantum gate in Ref.~\cite{QR_PRL}. In this section, we present the best known method to execute the local state preparation first introduced in Ref.~\cite{pitkapaperi}.

Our aim is to build a fixed structure circuit which takes any given input state $\ket{a}_n$ to any chosen state $\ket{b}_n$. We begin by noting that once we know an efficient circuit taking $\ket{a}_n$ to any fixed vector, for example $\ket{e_1}$, we may use the inverse that circuit with different parameters to transform $\ket{e_1}$ to $\ket{b}_n$. The $\ket{a}_n$ to $\ket{e_1}_n$ transformation consists of a sequence of gate pairs
\begin{equation}
S_a = \prod_{i=1}^n \left[ \left(\UC{i-1}{i}{R_y}\UC{i-1}{i}{R_z}\right) \otimes
I_{2^{n-i}} \right].
\end{equation}
The effect of the gate pair
$\UC{i-1}{i}{R_y}\UC{i-1}{i}{R_z}$
on the state $\ket{a}_i$ is to nullify half of its elements:
\begin{equation}
\UC{i-1}{i}{R_y}\UC{i-1}{i}{R_z} \ket{a}_i = \ket{a'}_{i-1}\otimes\ket{0}_1.
\end{equation}
Hence, each successive gate pair nullifies half of the elements of the state vector that have not yet been zeroed, and we have $S_a \ket{a}_n = \ket{e_1}_n$ up to a global phase.

\begin{figure}
\center
\includegraphics[width=0.80\textwidth]{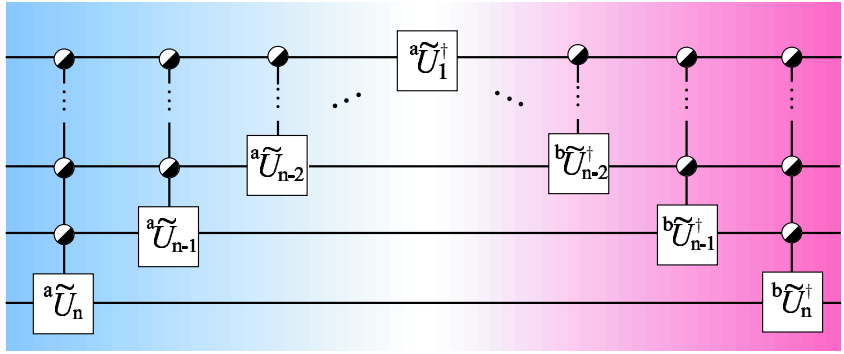}
\caption{\label{fig:vkaanto} Quantum circuit for transforming an arbitrary $n$-qubit state vector $\ket{a}_n$ into desired state vector $\ket{b}_n$. The resulting gates are of the form $\TUCU{i-1}{i}$ which is efficient to implement, see Fig.~\ref{fig:3tasaistaaskelta}.}
\end{figure}

Now we note that the pair of gates
$\UC{n-1}{n}{R_y}\UC{n-1}{n}{R_z} = \UCU{n-1}{n}$
may be replaced by the gate
\begin{equation}
\TUCU{n-1}{n}=\Delta_n^{\dagger} \UCU{n-1}{n},
\end{equation}
since the diagonal gate
\begin{equation}
\Delta_n^{\dagger} = \Delta_{n-1}^{0 \, \dagger}\otimes\ket{0}\bra{0}
+\Delta_{n-1}^{1 \, \dagger}\otimes\ket{1}\bra{1}
\end{equation}
does not mix the states;
\begin{align}
\Delta_n^{\dagger} \UCU{n-1}{n} \ket{a}_n &=
\Delta_n^{\dagger} \left(\ket{a'}_{n-1}\otimes\ket{0}_1\right) \notag \\
 &= \left(\Delta_{n-1}^{0 \, \dagger} \ket{a'}_{n-1}\right)\otimes\ket{0}_1 \notag \\
 &= \ket{a''}_{n-1}\otimes\ket{0}_1.
\end{align}
After combining $n-1$ pairs of adjacent $\UC{k}{k+1}{R_y}\UC{k}{k+1}{R_z}$ gates where $k=1, ..., n-1$ we find that the entire circuit for transforming $\ket{a}$ to $\ket{b}$ requires $2 \cdot 2^n -2n -2$ {\footnotesize{CNOT}}s and $2 \cdot 2^n -n -2$ one-qubit gates. If $\ket{a}$ or $\ket{b}$ coincides with one of the basis vectors $\ket{e_i}$, the gate counts are halved in the leading order. Figure~\ref{fig:vkaanto} showns the circuit diagram of the whole local state preparation $S_b^\dagger S_a$.

\section{Conclusion}\label{conclusion}
In this chapter we have studied efficient implementations of general $n$-qubit gates within the quantum circuit model. From the two philosophically different approaches, the cosine-sine decomposition based methods were found to lead to smaller gate counts than the QR decomposition based ones. The QR decomposition, the CSD and the NQ decomposition are compared in the required number of {\footnotesize{CNOT}}s and the total number of elementary gates in Tables.~\ref{tb:cnot} and~\ref{tb:total}, respectively. The QR decomposition is observed to have clearly the highest gate counts. The CSD requires slightly more {\footnotesize{CNOT}}s compared with the NQ decomposition but, on the other hand, the total number of elementary gates is noticeably larger in the NQ decomposition.

\begin{table}
\center
  \caption{\label{tb:cnot} Comparison of the number of {\footnotesize{CNOT}}s needed in different decompositions of general $n$-qubit gates.}
\begin{tabular}{|c|c|c|c|c|c|c|c|c|c|c|}
  \hline
   n   & 1 & 2 & 3 & 4 & 5 & 6 & 7 & 8 & 9 \\   \hline
  QR   & 0 & 4 & 64 & 536 & 4156 & 22618 & 108760 & 486052 & 2078668 \\
  CSD & 0 & 4  & 26  & 118 & 494  & 2014  & 8126   & 32638  &  130814 \\
  NQ  & 0 &  3 &  21 &  105 & 465  & 1953  & 8001  & 32385  & 130305 \\
  \hline
\end{tabular}
\end{table}

\begin{table}
\center
  \caption{\label{tb:total}   Comparison of the total number of gates needed in different decompositions of general $n$-qubit gates.}
\begin{tabular}{|c|c|c|c|c|c|c|c|c|c|c|}
  \hline
   n & 1 & 2 & 3 & 4 & 5 & 6 & 7 & 8 & 9 \\   \hline
  QR & 1 & 14 & 136 & 980 & 7384 & 42390 & 208820 & 944280 &  4062520 \\
  CSD & 1 & 11 & 58 & 249  & 1016 & 4087  & 16374  & 65525  &  262132 \\
  NQ   & 1 & 10  & 54  & 262 & 1142 & 4758  & 19414  & 78422  &  315222  \\
  \hline
\end{tabular}
\end{table}

A special class of gates, called uniformly controlled gates, was introduced as basic building blocks of quantum circuits. In fact, the power of the gate-efficient methods employing the cosine-sine decomposition lies deep on the efficient implementation of uniformly controlled rotations and two-qubit gates. These gates also proved to be essential in a circuit transforming an arbitrary quantum state into another, {\it i.e.}, performing local state preparation. In the case of a one-dimensional chain of qubits, the uniformly controlled one-qubit gates were decomposed using only nearest-neighbor gates, which may turn to be essential for the experimental realizations of quantum computers. By cleverly using the nearest-neighbor decomposition in the recursive CSD of an $n$-qubit gate, it has been shown~\cite{pitkapaperi} that only $\frac{5}{6} 4^n$ {\footnotesize{CNOT}}s are needed in the leading order. It is quite surprising that the gate count is increased by a factor of less than two, if the restriction to nearest-neighbor interactions is added.

In conclusion, we have reviewed the development of the circuit constructions of arbitrary quantum gates, slightly improved the lowest known gate count for the {\footnotesize{CNOT}}s to $\frac{23}{48}4^n$ in the leading order, discussed the local state preparation, and the circuits employing only nearest-neighbor {\footnotesize{CNOT}}s. 

\bibliography{kirja}

\begin{thebibliography}{10}

\bibitem{Ballentine}
Ballentine, L.~E. {\em Quantum Mechanics: a Modern Development}; World Scientific, Singapore, 1998.

\bibitem{NielsenChuang}
Nielsen, M.~A.; Chuang, I.~L. {\em Quantum Computation and Quantum
  Information}; Cambridge University Press: Cambridge, 2000.

\bibitem{shor94}
Shor, P.~W. {\em{IEEE} Proc. 35nd Annual Symposium on Foundations of Computer
  Science} 1994, 124.

\bibitem{gisin}
Gisin, N.; Ribordy, G.; Tittel, W.; Zbinden, H. {\em Rev. Mod. Phys.} 2002, vol. 74, 145.

\bibitem{deutsch}
Deutsch, D. {\em Proc. R. Soc. of Lond. A} 1989, vol. 425,  73.

\bibitem{barenco}
Barenco, A.; Bennett, C.~H.; Cleve, R.; DiVincenzo, D.~P.; Margolus, N.~H.;
  Shor, P.~W.; Sleator, T.; Smolin, J.~A.; Weinfurter, H. {\em Phys. Rev. A} 1995, vol. 52,  3457.

\bibitem{lloyd}
Lloyd, S. Phys. {\em Rev. Lett.} 1995, vol. 75,  346.

\bibitem{knill}
Knill, E.; quant-ph/9508006,  1995.

\bibitem{cybenko}
Cybenko, G. {\em Computing in Science and Engineering} 2001, vol. 3,  27.

\bibitem{shende}
Shende, V.~V.; Markov, I.~L.; Bullock, S.~S. {\em Phys. Rev. A} 2004, vol. 69,
  062321.

\bibitem{QR_PRL}
Vartiainen, J.~J.; M\"ott\"onen, M.; Salomaa, M.~M. {\em Phys. Rev. Lett.} 2004, vol. 92,  177902.

\bibitem{golub}
Golub, G.~H.; Van~Loan, C.~F. {\em Matrix Computations} 3rd ed.; Johns Hopkins Press: Baltimore, 1996.

\bibitem{CSD_PRL}
M\"ott\"onen, M.; Vartiainen, J.~J.; Bergholm, V.; Salomaa, M.~M.
  {\em Phys. Rev. Lett.} 2004, vol. 93,  130502.

\bibitem{tucci}
Tucci, R.~R.; quant-ph/9902062,  1999.

\bibitem{pitkapaperi}
Bergholm, V.; Vartiainen, J.~J.; M\"ott\"onen, M.; Salomaa, M.~M.; quant-ph/0410066, 2004.

\bibitem{shende_matrix}
Shende, V.~V.; Bullock, S.~S.; Markov, I.~L.; quant-ph/0406176.

\bibitem{stateprep}
M\"ott\"onen, M.; Vartiainen, J.~J.; Bergholm, V.; Salomaa, M.~M.; quant-ph/0407010,  2004.

\bibitem{tucci2}
Tucci, R.~R.; quant-ph/0411027,  2004.

\bibitem{bullock_quditl}
Bullock, S.~S.; Brennen, G.~K.; O'Leary, D.~P.; quant-ph/0410116,  2004.

\bibitem{bullock:2004}
Bullock, S.~S.; Markov, I.~L. {\em Quant. Inf. and Comp.} 2004, vol. 4,  27.

\bibitem{tucci3}
Tucci, R.~R.; quant-ph/0407215,  2004.

\bibitem{Givens}
Givens, W. J. {\em Soc. Ind. Appl. Math} 1958, vol. 6,  26.

\bibitem{aho}
Aho, A.~V.; Svore, K.~M.; quant-ph/0322008,  2003.

\bibitem{oleary}
O'Leary, D.~P.; Bullock, S.~S.; unpublished,  2004.

\bibitem{Paige}
Paige, C.~C.; Wei, M. {\em Linear Algebra and Appl.} 1994, vol. 208,  303.

\bibitem{ruler}
Guy, R.~K., {\em Unsolved Problems in Number Theory, 2nd ed.}; Springer-Verlag:
  New York, 1994; p.\ 224.

\bibitem{Shende_vector}
Shende, V.~V.; Markov, I.~L.; quant-ph/0401162,  2004.

\end{thebibliography}
\end{document}